# New Nanoporous Graphyne Monolayer as Nodal Line Semimetal: Double Dirac Points with an Ultrahigh Fermi Velocity

Linyang Li[†,*], Xiangru Kong[‡,*], François M. Peeters[†,*]

[†]*Department of Physics, University of Antwerp, Groenenborgerlaan 171, B-2020 Antwerp, Belgium*

[‡]*International Center for Quantum Materials, Peking University, 100871 Beijing, China*

**Abstract:** Two-dimensional (2D) carbon materials play an important role in nanomaterials. We propose a new carbon monolayer, named hexagonal-4,4,4-graphyne ($H_{4,4,4}$-graphyne), which is a nanoporous structure composed of rectangular carbon rings and triple bonds of carbon. Using first-principles calculations, we systematically studied the structure, stability, and band structure of this new material. We found that its energy is much lower than that of some experimental carbon materials and it is stable at least up to 1500 K. In contrast to the single Dirac point band structure of other 2D carbon monolayers, the band structure of $H_{4,4,4}$-graphyne exhibits double Dirac points along the high symmetry points and the corresponding Fermi velocities (1.04~1.27 $\times$ 10$^6$ m/s) are asymmetric and higher than that of graphene. The origin of these double Dirac points is traced back to the nodal line states, which can be well explained by a tight-binding model. The $H_{4,4,4}$-graphyne forms a moiré superstructure when placed on top of a BN substrate, while keeping the double Dirac points. These properties make $H_{4,4,4}$-graphyne a promising semimetal material for applications in high-speed electronic devices.

*Corresponding authors. E-mail addresses:
linyang.li@uantwerpen.be (Linyang Li)
kongxru@pku.edu.cn (Xiangru Kong)
francois.peeters@uantwerpen.be (François M. Peeters)

**Introduction**

Monolayer graphene was first realized in 2004[1] and since then two-dimensional (2D) carbon material research has played a crucial role in nanomaterials. Many kinds of 2D carbon allotropes have been proposed due to the huge flexibility of the carbon bonding. Graphdiyne,[2] a special structure of graphyne, has been realized experimentally. Topological defects,[3-5] which include non-hexagonal carbon rings, have been observed in graphene. Many more new 2D carbon allotropes have been predicted theoretically with novel crystal structures that can be classified into two general classes. The first are the carbon monolayers that include some non-hexagonal carbon rings,[6-8] such as haeckelite $H_{5,6,7}$[9]/phagraphene[10]/$\Psi$-graphene[11] (5-6-7 rings), $T$ graphene[12] (4-8 rings), and penta-graphene[13] (5 rings). These structures exhibit $sp^2/sp^3$ hybridization of the carbon atom. The second one are the carbon monolayers that have the triple bonds of carbon (-C≡C-) due to the $sp$ hybridization of the carbon atom, such as $\alpha/\beta/\gamma/\delta$/6,6,12-graphyne,[14-16] in which the carbon atoms ($sp^2$ hybridization)/hexagonal carbon rings are connected by -C≡C-. These 2D carbon allotropes show different fundamental physical and chemical properties. Not only their band structures change from metal/semimetal to semiconductor, but also they can be used in many energetic and environmental applications, such as for gas separation,[17] and for water desalination.[18] The abundant new 2D carbon structures also provide efficient inspiration for structural predictions of other elements,[19] leading to many more new lattice structures with excellent properties. Although many 2D carbon allotropes have been predicted, combining structural properties of the above two classes have been scarce up to now.[20-22]

The topologically nontrivial materials, such as topological semimetals, have attracted broad interest. There are three distinct kinds of topological semimetals: Dirac, Weyl, and nodal line semimetals.[23] For the nodal line semimetals, the band crossing points form a continuous Dirac loop with a relatively higher density of states at the Fermi level,[24] which is an advantage for high-speed electronic devices. Many kinds of three-dimensional (3D) nodal line bulk materials, such as $PtSn_4$,[25] $PbTaSe_2$,[26] and ZrSiS,[27,28] have been realized experimentally. Theoretically, $TlTaSe_2$,[29] 3D-honeycomb graphene networks,[30] $Ca_3P_2$,[31,32] LaN,[33] $Cu_3PdN$,[34] and body-centered orthorhombic $C_{16}$,[35] have been predicted to show nodal line states.[36] In contrast to the extended literature on 3D nodal line semimetals, the study of semimetal nodal line states in 2D materials are still in its infancy. They have only been confirmed experimentally in $Cu_2Si$ monolayer[37] and theoretical predictions of a nodal line band structure have been made for a few 2D materials, such as $Be_2C/BeH_2$,[38] MX (M = Pd, Pt; X = S, Se, Te),[24] and $A_3B_2$ compound (A is a group-IIB cation and B is a group-VA anion, such as $Hg_3As_2$).[39] Therefore, there is a need for more predictions of new 2D nodal line semimetals that are stable and that can be realized experimentally.

In this work, we constructed a new graphyne monolayer with a hexagonal lattice using rectangular carbon rings and triple bonds of carbon. According to the naming rule of graphyne and its lattice feature, $H_{4,4,4}$-graphyne is obtained. Using first-principles calculations, we systematically investigated the structure, energy, stability, and electronic band structure of the $H_{4,4,4}$-graphyne monolayer. This monolayer shows a nanoporous structure and its total energy is almost equal to that of $β$-graphyne. The phonon spectrum provides convincing evidence for the dynamical stability of $H_{4,4,4}$-graphyne and our molecular dynamics (MD) calculations show that the monolayer is stable up to a high temperature. Different to the band structure with a single Dirac point of most other carbon monolayers, $H_{4,4,4}$-graphyne has a band structure of double Dirac points along the high symmetry points with high Fermi velocities, which we confirm using different calculation methods. By an analysis of the orbital-projected band structure, $p_z$ atomic orbitals of the carbon atoms are responsible for the double Dirac points in the $H_{4,4,4}$-graphyne monolayer. Using the $p_z$ atomic orbitals, a tight-binding (TB) model is constructed, which not only reproduces the double Dirac points, but also shows that the physical origin of the double Dirac points can be traced back to the nodal line states. Finally, we show that the $H_{4,4,4}$-graphyne/BN moiré superstructure is a possible way of realizing the $H_{4,4,4}$-graphyne monolayer experimentally.

**Method**

Our first-principles calculations were performed using the Vienna *ab initio* simulation package (VASP) code,[40-42] implementing density functional theory (DFT). The electron exchange-correlation functional was treated by using the generalized gradient approximation in the form proposed by Perdew, Burke, and Ernzerhof (PBE).[43] The atomic positions and lattice vectors were fully optimized using the conjugate gradient scheme until the maximum force on each atom was less than 0.01 eV/Å. The energy cutoff of the plane-wave basis was set to 520 eV with an energy precision of $10^{-5}$ eV. The Brillouin zone (BZ) was sampled by using a $9 \times 9 \times 1$ Γ-centered Monkhorst-Pack grid. The vacuum space was set to at least 15 Å in all the calculations to minimize artificial interactions between neighboring slabs. The phonon spectrum was calculated using a supercell ($4 \times 4$) approach within the PHONOPY code.[44]

**Structure**

The investigated graphyne monolayer is shown in Figure 1(a). Its hexagonal framework structure is composed of rectangular carbon rings, connected by the triple bonds of carbon. The carbon atoms in the rectangular carbon rings are close to $sp^2$ hybridization, because the angle of the two neighboring single bonds of carbon (-C−C-) is not equal to 120° while the carbon atoms in the -C≡C- are close to $sp$ hybridization, because the four carbon atoms (-C−C≡C−C-) are not located in a strict straight line.

According to the naming rule of graphyne, the new graphyne should be named as 4,4,4-graphyne.[45] To distinguish the rectangular graphyne (R-graphyne),[46] which can also be called as 4,4,4-graphyne using the naming rule of graphyne, we call our proposed graphyne hexagonal-4,4,4-graphyne ($H_{4,4,4}$-graphyne). Most new predicted graphyne structures are based on the hexagonal graphene structure by inserting the triple bonds of carbon, such as 6,6,12-graphyne,[16] 14,14,14-graphyne,[47] and α-graphyne (α-2/α-3/α-4 graphyne),[48,49] while those with non-hexagonal carbon rings are rather exceptional.[20-22,46] Here, we provide a novel structure model in which the rectangular carbon rings and triple bonds of carbon can coexist, providing a novel structure model for new stable carbon monolayers.

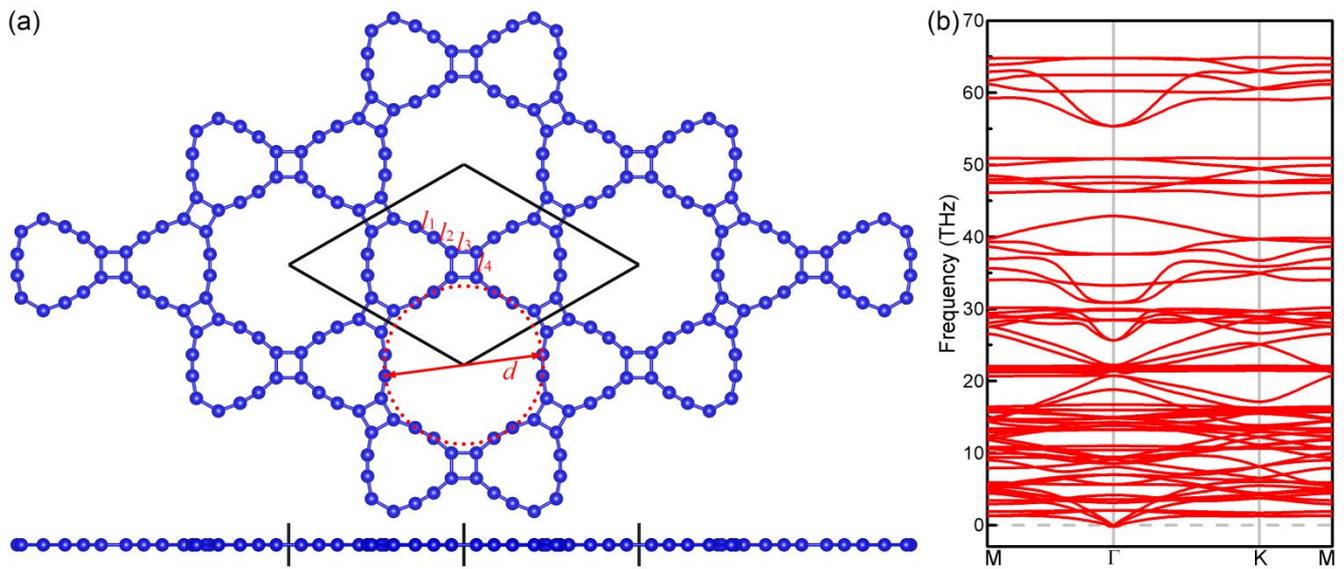

Figure 1. Schematic representations (top and side views) of $H_{4,4,4}$-graphyne (a) and its phonon spectrum along the high-symmetry points in the BZ (b). The blue dots are the carbon atoms. The four kinds of bond lengths are labeled $l_1$, $l_2$, $l_3$, and $l_4$, and the diameter of the circumcircle (red dotted line) of the 24 carbon ring is labeled $d$. The black box is the unit cell.

The lattice of the $H_{4,4,4}$-graphyne is not only a hexagonal lattice, but also a kagome lattice,[39] which is formed by all the centers of the rectangular carbon rings. Its lattice constant is 11.82 Å. There are four kinds of bond lengths, which are labeled $l_1$, $l_2$, $l_3$, and $l_4$, as shown in Figure 1(a). The $l_1$ = 1.247 Å is the length of the carbon triple bond, which is formed by two carbon atoms close to *sp* hybridization and the value of $l_1$ is close to that of the other graphyne structures.[14] All the other carbon atoms are close to $sp^2$ hybridization with bond lengths, $l_2$ = 1.351 Å, $l_3$ = 1.453 Å, and $l_4$ = 1.489 Å. Since H$_{4,4,4}$-graphyne can be regarded as inserting -C≡C- into the -C−C- of graphenylene that are shared by 6 and 12 carbon rings,[17] the graphenylene bond lengths (1.367/1.474/1.473 Å) are close to those of our structure ($l_2$/$l_3$/$l_4$). The nonoporous graphenylene membrane has been theoretically predicted to achieve efficient $^3$He/$^4$He separation for industrial applications,[17] while the graphyne membrane has been proposed for water desalination.[18] Besides these carbon monolayers, the C$_2$N-*h*2D[50] and g-C$_3$N$_4$[51] membranes have also been proposed for separation applications, depending on their nonoporous structure. The diameter $d$ = 9.30 Å of the circumcircle (red dotted line in Figure 1(a)) of the 24 carbon ring of H$_{4,4,4}$-graphyne is much larger than the one of other monolayers (5.49 Å for graphenylene,[17] 6.90 Å for graphyne-3,[18] 5.51 Å for C$_2$N-*h*2D,[50] and 4.76 Å for g-C$_3$N$_4$[51]) and is comparable to that of the well-known covalent organic frameworks, which have been shown to have potential for a wide range of applications in gas/liquid separation.[52]

**Energy and Stability**

Introducing *sp*-hybridized carbon atoms can increase the system total energy ($E_t$ with a unit of eV per carbon atom) of carbon allotropes. Graphene shows the lowest $E_t$ in all the carbon allotropes while the graphyne structures have a higher $E_t$ due to the presence of *sp*-hybridized carbon atoms.[15] Setting $n$ ($N$) as the number of *sp*-hybridized carbon atoms (total carbon atoms) in the unit cell, we calculated the ratio $n/N$. For $\gamma$-graphyne, $\beta$-graphyne, and $\alpha$-graphyne, $E_t$ increases with the value of this ratio (Table 1). However, H$_{4,4,4}$-graphyne shows the same ratio (0.50) as $\gamma$-graphyne, but its $E_t$ is higher than that of $\gamma$-graphyne and comparable to that of $\beta$-graphyne. This can be ascribed to the special structure of H$_{4,4,4}$-graphyne. In contrast to fully $sp^2$-hybridized carbon atoms of $\gamma$-graphyne in which the three angles of the two neighboring single bonds are all 120°,[14] the three angles in our structure are 140.1° (belonging to the 24 carbon ring), 129.9° (belonging to the 12 carbon ring), and 90° (belonging to the 4 carbon ring), which leads to an increase of $E_t$. The similar situations can apply to the $sp$-hybridized carbon atoms, where the angle between single bond and triple bond of H$_{4,4,4}$-graphyne becomes 170.1° (belonging to the 12 carbon ring) instead of 180° of $\gamma$-graphyne. Although H$_{4,4,4}$-graphyne shows a higher $E_t$ as compared to graphene, it is still energetically preferable over some experimentally investigated carbon nanostructures, such as the C$_{20}$ fullerene[13] and the T-carbon nanowire.[53] The calculated $E_t$ of T-carbon is -7.92 eV/atom,[54] which is less favorable than the -8.37 eV/atom of the proposed H$_{4,4,4}$-graphyne structure, implying that there is a good chance that H$_{4,4,4}$-graphyne can be synthesized in the future.

| 2D carbon allotropes | Ratio ($n/N$) | $E_t$ (eV/atom) |
|---|---|---|
| Graphene | 0 (0/2) | -9.22 |
| H$_{4,4,4}$-graphyne | 0.50 (12/24) | -8.37 |
| $\gamma$-graphyne | 0.50 (6/12) | -8.58 |
| $\beta$-graphyne | 0.67 (12/18) | -8.38 |
| $\alpha$-graphyne | 0.75 (6/8) | -8.30 |

**Table 1.** Calculated ratio of $sp$-hybridized carbon atoms ($n$) to the total carbon atoms ($N$) in the unit cell and total energy $E_t$ (eV/atom) of different 2D carbon allotropes.

Next, we studied the stability of H$_{4,4,4}$-graphyne from dynamical and thermal aspects. The phonon spectrum of H$_{4,4,4}$-graphyne is shown in Figure 1(b). The phonon spectrum is free from imaginary frequency modes, which indicates that the $H_{4,4,4}$-graphyne monolayer is dynamically stable. Then we confirm the thermal stability of the $H_{4,4,4}$-graphyne monolayer by first-principles MD simulations. We used a 2×2 supercell to perform MD simulations at 500K, 1000K and 1500K for 10 ps with a time step of 1 fs. The fluctuations of the total energy with time at the three temperatures are shown in Figures 2(a), (b), and (c), and the corresponding snapshot of the atomic configuration after the MD simulations (10 ps) is given at the bottom of Figure 2(a)/(b)/(c). The total energy of the system converges within this time scale. The final geometrical framework of the $H_{4,4,4}$-graphyne structure containing the 4-12-24 carbon rings is well preserved and no structure reconstruction is found to occur in all the three cases. The three different temperatures have little influence on the nanoporous structure, implying that $H_{4,4,4}$-graphyne monolayer is robust. The above dynamical and thermal results indicate that the $H_{4,4,4}$-graphyne monolayer is stable at least up to 1500K, which shows its great potential for applications in a high temperature environment.

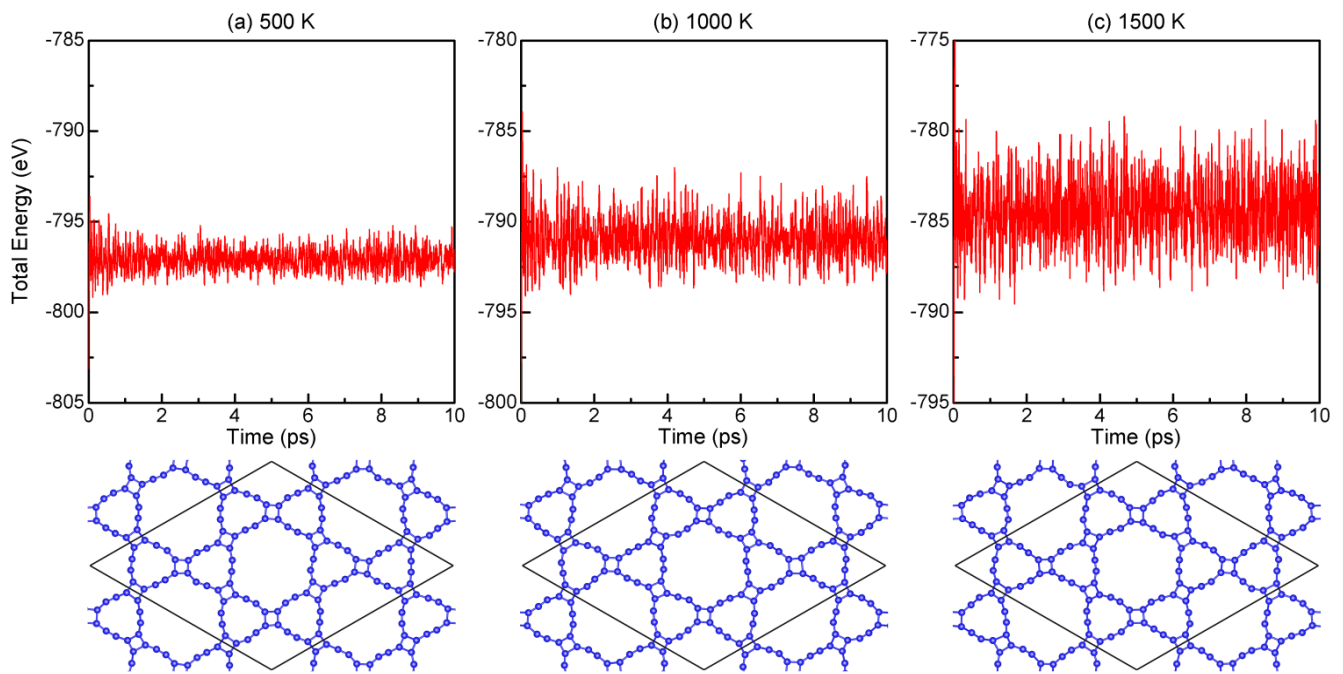

Figure 2. Fluctuations of total energy as function of simulation time and the corresponding snapshots of the atomic configuration (top view) after the MD simulations (10 ps) at the temperature of 500 K (a), 1000K (b), and 1500 K (c).

**Band Structure**

The electronic band structure of the $H_{4,4,4}$-graphyne monolayer is shown in Figure 3. The red lines indicate the band structure at the PBE level (Figure 3(a)), which shows two Dirac points at the P (M-Γ) and Q (Γ-K) points along the high-symmetry points in the BZ. An enlarged view of the bands at the P/Q point near the Fermi level is presented in Figure 4(a)/(b), which shows that two bands cross linearly at the Fermi level and thus the charge carriers can be characterized by massless Dirac fermions. To fully confirm the existence of the double Dirac points, we used the more sophisticated Heyd-Scuseria-Ernzerhof (HSE06)[55,56] hybrid functional method to calculate the band structure of $H_{4,4,4}$-graphyne monolayer, which is shown in Figure 3(b) (blue lines). Similar double Dirac points at P' (M-Γ) and Q' (Γ-K) points can also be clearly seen. From the enlarged view of the bands at the P'/Q' point near the Fermi level (Figure 4(c)/(d)), we confirm that the two bands cross linearly at the Fermi level, which is similar to the result of PBE. Although the PBE method typically underestimates the band gap in semiconductors, the fact that the HSE06 method gives the same Dirac points[57,58] strengthens us in the validity of the band structure around the Fermi level. A few 2D carbon structures exhibit a Dirac point at the K point with a high velocity, such as α-graphyne,[14] δ-graphyne,[15] and graphene. Phagraphene[10] and β-graphyne[59] have a distorted Dirac point, which is not located in one of the high-symmetry points. In contrast to the above band structures with a single Dirac point, 6,6,12-graphyne[16] and buckled T graphene[12] show a band structure with double Dirac points along the high symmetry points. Although $H_{4,4,4}$-graphyne and buckled T graphene have a similar double Dirac points around the Γ point, there is a major difference. The two Dirac points of $H_{4,4,4}$-graphyne are at the same energy ($|E(P) - E(Q)| < 0.05$ meV and $|E(P') - E(Q')| < 0.09$ meV), which is different from that of buckled $T$ graphene where the two Dirac points are separated by an energy of 25 meV.[12]

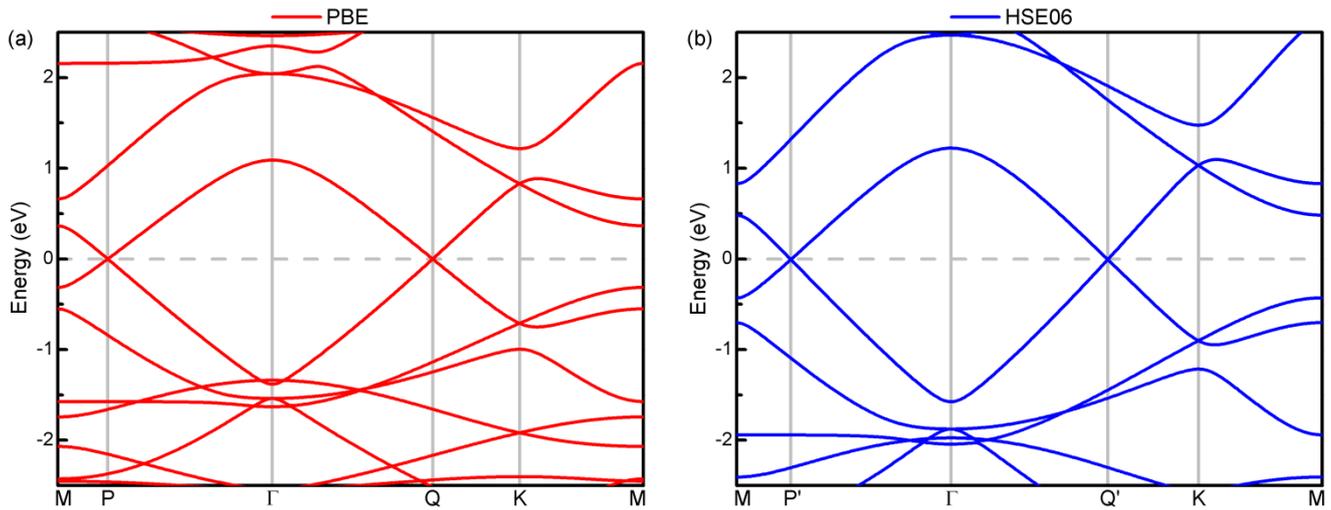

Figure 3. Band structures of the $H_{4,4,4}$-graphyne monolayer from PBE calculations (red lines, (a)) and HSE06 calculations (blue lines, (b)). The Dirac point in reciprocal space is labeled as P/Q/P'/Q'. The energy at the Fermi level was set to zero.

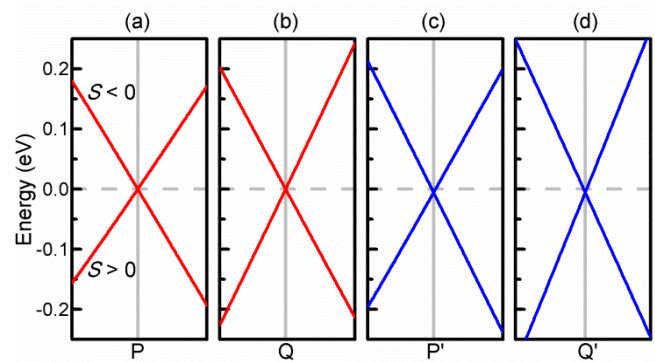

Figure 4. An enlarged view of the bands at the P (a)/Q (b)/P'(c)/Q'(d) point near the Fermi level along the direction of M-Γ/Γ-K/M-Γ/Γ-K. Corresponding to Figure 3, the red/blue lines are the band structures from PBE/HSE06 calculations. The slope of the band close to the Dirac point P is indicated by $S$ and a similar label can be applied to Q/P'/Q'.

From the Dirac point with linear bands, we can calculate the Fermi velocity ($v_F$) by a linear fitting of the first-principles calculations date. Because the Dirac points are not at the high symmetry points, the double Dirac points can be regarded as two distorted Dirac points, which is similar to that in phagraphene[10]/distorted GaBi-X$_2$ monolayers (X = I, Br, Cl)[60]/YN$_2$.[61] In general, the distorted Dirac point has two different Fermi velocities along the high-symmetry line directions. Setting the slope of the bands close to the Dirac point along the M-Γ(Γ-K) direction as $S$ (Figure 4(a)), we obtain two kinds of slopes ($S > 0$ and $S < 0$) at each Dirac point, and thus two different Fermi velocities can be obtained. For the band structures at the PBE level (red lines, Figures 4(a) and (b)), the Fermi velocities of the two distorted Dirac points are $v_F$ (P, $S > 0$) = 0.87 $\times 10^6$ m/s, $v_F$ (P, $S < 0$) = 0.99 $\times 10^6$ m/s, $v_F$ (Q, $S > 0$) = 1.07 $\times 10^6$ m/s, and $v_F$ (Q, $S < 0$) = 0.95 $\times 10^6$ m/s. Comparing these results with that of graphene at the PBE level, $v_F$ (K) = 0.83 $\times 10^6$ m/s, the Fermi velocities of H$_{4,4,4}$-graphyne (0.87~1.07 $\times 10^6$ m/s) are slightly high. It is well known that there are many kinds of 2D carbon structures with a Dirac point, but their Fermi velocities are all lower than that of graphene,[10] such as $\alpha$-graphyne (0.687 $\times 10^6$ m/s)[49] and $\delta$-graphyne (0.696 $\times 10^6$ m/s).[15] To our knowledge, the Fermi velocities of H$_{4,4,4}$-graphyne are the highest Fermi velocities among all the predicted 2D carbon structures. To further confirm the superiority to graphene, we also calculated the four Fermi velocities at the HSE06 level (blue lines, Figure 4(c) and (d)). The Fermi velocities are $v_F$ (P', $S > 0$) = 1.04 $\times 10^6$ m/s, $v_F$ (P', $S < 0$) = 1.19 $\times 10^6$ m/s, $v_F$ (Q', $S > 0$) = 1.27 $\times 10^6$ m/s, and $v_F$ (Q', $S < 0$) = 1.13 $\times 10^6$ m/s. These results (1.04~1.27 $\times 10^6$ m/s) are also slightly higher than the result for graphene at the HSE06 level, $v_F$ (K) = 1.01 $\times 10^6$ m/s. From the above results obtained within different calculations methods, we confirm the ultrahigh Fermi velocity of

$H_{4,4,4}$-graphyne, which is advantageous for building high-speed electronic devices, such as field effect transistor.

To investigate the origin of the double Dirac points, we calculated the orbital-projected band structure along the high-symmetry points in the BZ at the PBE level as shown in Figure 5(a). It is clear that the bands (red dots) including the double Dirac points close to the Fermi level originate from the $p_z$ atomic orbitals of the carbon atoms. This is similar to most other 2D carbon structures having a Dirac point, such as phagraphene,[10] $\delta$-graphyne,[15] and $\alpha$-graphyne[49] and the interaction between the $p_z$ atomic orbitals leads to the formation of a π-conjugated framework. The $s+p_x+p_y$ atomic orbitals only contribute to the bands (blue dots) located in the region around 1.2 eV (valence band) and 2.4 eV (conduction band) away from the Fermi level.

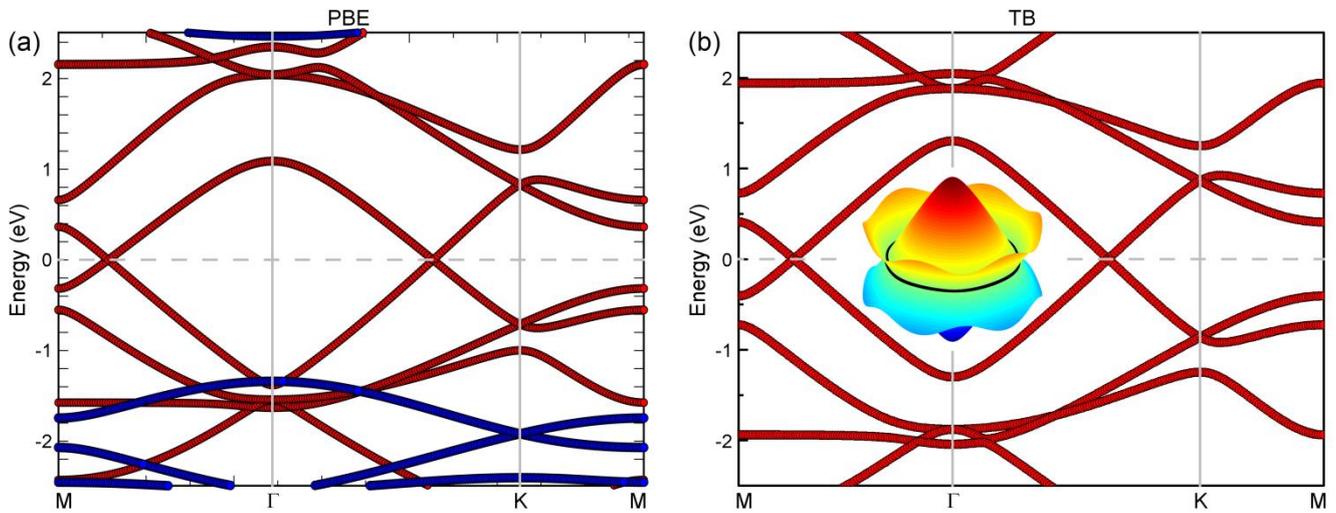

Figure 5. Orbital-projected band structure of $H_{4,4,4}$-graphyne monolayer from PBE calculations (a) and band structure of $H_{4,4,4}$-graphyne monolayer from TB calculations (b). In figure (a), the red (blue) dots represent the contributions from the $p_z$ ($s+p_x+p_y$) atomic orbitals of the carbon atoms. The red dots in figure (b) correspond to those in figure (a). The 3D band structure from TB calculations around the nodal line is presented as an insert in the figure (b).

For most 2D carbon structures, their band structures can be well reproduced by a TB model. To further illustrate the band structure with the double Dirac points of $H_{4,4,4}$-graphyne monolayer, we propose a TB model involving only the $p_z$ atomic orbitals of the twenty-four carbon atoms in the unit cell. The effective Hamiltonian is taken as[15]

$$H = -\sum_{\langle ij \rangle}(t_{ij}c_i^+ c_j + h.c.) ,$$

where $c_i^+$ and $c_i$ represent the creation and annihilation operators of an electron at the $i$-th atom, respectively. Since all atoms are carbon atoms, we can neglect the on-site energy difference of these carbon atoms and set it as zero. To get a better understanding for the appearance of the double Dirac points, we only consider the hopping of the $p_z$ atomic orbitals between the nearest-neighboring atoms. Corresponding to the four bond lengths ($l_1$, $l_2$, $l_3$, and $l_4$), there are four hopping parameters $t(l_1)$, $t(l_2)$, $t(l_3)$, and $t(l_4)$. The distance-dependent hopping energy is determined by the formula $t(l_m) = t_0 \times \exp(q \times (1-l_m/l_0))$,[10] where $t_0 = 2.7$ eV, $q = 2.2$, $l_0 = 1.5$ Å, and $m = 1, 2, 3, 4$. We can obtain the band structure of $H_{4,4,4}$-graphyne by diagonalizing a 24×24 matrix in reciprocal space and the result is shown in Figure 5(b). Corresponding the red dots representing the contributions from the $p_z$ atomic orbitals of the carbon atoms in Figure 5(a), the TB bands that are also indicated by the red dots are in good agreement with the PBE results; in particular, the double Dirac points at the Fermi level are very accurately reproduced.

For the double Dirac points in 2D carbon structures, there are two kinds of origins. One is the double Dirac cones as shown in the Cp-graphyne[20] and 6,6,12-graphyne,[59] where the two Dirac points come from the two Dirac cones. Another one is that the linear dispersion relation near the Fermi level exists in each direction forming a Dirac loop, which is called a nodal line band structure, such as in buckled *T* graphene.[12] To further distinguish the two kinds of origins in $H_{4,4,4}$-graphyne, we also calculated the 3D band structure (insert of Figure 5(b)) around the double Dirac points. Notice that the two band lines forming the double Dirac points become two band surfaces, which cross at the Fermi level forming a Dirac loop (black line in insert of Figure 5(b)). This kind of band structure should obviously be a Dirac nodal line band structure, which implies that $H_{4,4,4}$-graphyne is a 2D Dirac nodal line semimetal.[24,37-39,62]

**Moiré Superstructure**

BN is an appealing substrate material, because it has an atomically smooth surface that is relatively free of dangling bonds and charge traps.[63] It has been a standard substrate for graphene, as confirmed experimentally and theoretically.[64,65] Different from the simple stacking models, such as AA and AB, the graphene/BN heterostructure results in a moiré superstructure stacking model and the interaction between the two planer layers is due to van der Waals (vdW) force.[64,65] In theoretical calculations and experimental synthesis, since the BN substrate can well preserve the hexagonal honeycomb structure of graphene,[64,65] BN has been a preferential substrate for other 2D monolayer structures, such as silicene[66] and germanene.[67] Here, we constructed a $H_{4,4,4}$-graphyne/BN heterostructure with the moiré superstructure stacking model. The superstructure is shown in Figure 6(a), in which we used the $\sqrt{21} \times \sqrt{21}$ BN supercell to match the 1×1 $H_{4,4,4}$-graphyne. In the PBE calculations of the superstructure, the vdW interaction is included (DFT-D3).[68] Similar to other moiré superstructures, such as graphene/BN,[65] silicene/BN,[66] germanene/BN,[67] and silicene/MoS$_2$,[69] an obviously rotation angle between the lattices of $H_{4,4,4}$-graphyne and BN substrate can be seen. The corresponding band structure is shown in Figure 6(b). It is clear that the bands from the BN substrate are far away from the Femi level and the band structure with double Dirac points is well preserved, which is similar to the case of graphene/BN.[65] We therefore propose that the BN substrate may be an ideal substrate for $H_{4,4,4}$-graphyne, which contributes to the stabilization of the monolayer and the preservation of the double Dirac points.

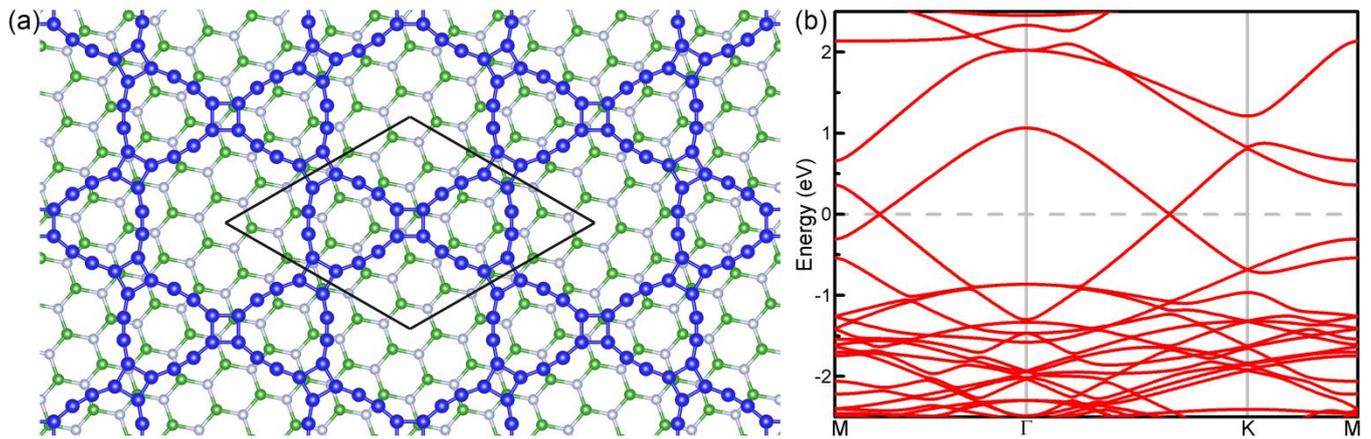

Figure 6. Schematic representation (top view) of $H_{4,4,4}$-graphyne/BN moiré superstructure (a) and corresponding band structure from PBE calculations (b). The blue/green/silver dots are carbon/boron/nitride atoms.

**Conclusion**

In summary, using first-principles calculations combined with a TB model, we predict that the new carbon monolayer, $H_{4,4,4}$-graphyne, is a nodal line semimetal with: 1) a nanoporous structure, 2) high stability, 3) band structure with double Dirac points, and 4) ultrahigh Fermi velocities. The nanoporous structure shows potential applications for gas/liquid separation. The value of the total energy, phonon calculations, and MD simulations fully confirm its energetic, dynamical, and thermal stability. The double Dirac points with ultrahigh Fermi velocities make $H_{4,4,4}$-graphyne a promising material for high-speed electronic devices. A simple TB model was constructed and it was showed that the origin of the double Dirac points is the nodal line states. The $H_{4,4,4}$-graphyne/BN moiré superstructure shows a possible way of realizing H$_{4,4,4}$-graphyne monolayer experimentally, which preserves its double Dirac points simultaneously.

**Acknowledgements**

This work was supported by the Fonds voor Wetenschappelijk Onderzoek (FWO-Vl), and the FLAG-ERA project TRANS2DTMD. The computational resources and services used in this work were provided by the VSC (Flemish Supercomputer Center), funded by the Research Foundation - Flanders (FWO) and the Flemish Government – department EWI.

**References**

(1) Novoselov, K. S.; Geim, A. K.; Morozov, S. V.; Jiang, D.; Zhang, Y.; Dubonos, S. V.; Grigorieva, I. V.; Firsov, A. A. Electric Field Effect in Atomically Thin Carbon Films. *Science* **2004**, *306*, 666–669.

(2) Li, G.; Li, Y.; Liu, H.; Guo, Y.; Li, Y.; Zhu, D. Architecture of graphdiyne nanoscale films. *Chem. Commun.* **2010**, *46*, 3256–3258.

(3) Yazyev, O. V.; Chen, Y. P. Polycrystalline graphene and other two-dimensional materials. *Nat. Nanotechnol.* **2014**, *9*, 755–767.

(4) Lahiri, J.; Lin, Y.; Bozkurt, P.; Oleynik, I. I.; Batzill, M. An extended defect in graphene as a metallic wire. *Nat. Nanotechnol.* **2010**, *5*, 326–329.

(5) Cockayne, E.; Rutter, G. M.; Guisinger, N. P.; Crain, J. N.; First, P. N.; Stroscio, J. A. Grain boundary loops in graphene. *Phys. Rev. B* **2011**, *83*, 195425.

(6) Zheng, G.; Jia, Y.; Gao, S.; Ke, S. H. A planar carbon allotrope with linear bipentagon-octagon and hexagon arrangement. *Physica E* **2017**, *87*, 107–111.

(7) Sharma, B. R.; Manjanath, A.; Singh, A. K. *pentahexoctite*: A new two-dimensional allotrope of

carbon. *Sci. Rep.* **2014**, *4*, 7164.

(8) Sun, H.; Mukherjee, S.; Daly, M.; Krishnan, A.; Karigerasi, M. H.; Singh, C. V. New insights into the structure-nonlinear mechanical property relations for graphene allotropes. *Carbon* **2016**, *110*, 443–457.

(9) Terrones, H.; Terrones, M.; Hernández, E.; Grobert, N.; Charlier, J. C.; Ajayan, P. M. New Metallic Allotropes of Planar and Tubular Carbon. *Phys. Rev. Lett.* **2000**, *84*, 1716–1719.

(10) Wang, Z.; Zhou, X. F.; Zhang, X.; Zhu, Q.; Dong, H.; Zhao, M.; Oganov, A. R. Phagraphene: A Low-Energy Graphene Allotrope Composed of 5−6−7 Carbon Rings with Distorted Dirac Cones. *Nano Lett.* **2015**, *15*, 6182–6186.

(11) Li, X.; Wang, Q.; Jena, P. ψ-Graphene: A New Metallic Allotrope of Planar Carbon with Potential Applications as Anode Materials for Lithium-Ion Batteries. *J. Phys. Chem. Lett.* **2017**, *8*, 3234–3241.

(12) Liu, Y.; Wang, G.; Huang, Q.; Guo, L.; Chen, X. Structural and Electronic Properties of *T* Graphene: A Two-Dimensional Carbon Allotrope with Tetrarings. *Phys. Rev. Lett.* **2012**, *108*, 225505.

(13) Zhang, S.; Zhou, J.; Wang, Q.; Chen, X.; Kawazoe, Y.; Jena, P. Penta-graphene: A new carbon allotrope. *Proc. Natl. Acad. Sci.* **2015**, *112*, 2372–2377.

(14) Kim, B. G.; Choi, H. J. Graphyne: Hexagonal network of carbon with versatile Dirac cones. *Phys. Rev. B* **2012**, *86*, 115435.

(15) Zhao, M.; Dong, W.; Wang, A. Two-dimensional carbon topological insulators superior to graphene. *Sci. Rep.* **2013**, *3*, 3532.

(16) Malko, D.; Neiss, C.; Görling, A. Two-dimensional materials with Dirac cones: Graphynes containing heteroatoms. *Phys. Rev. B* **2012**, *86*, 045443.

(17) Qu, Y.; Li, F.; Zhao, M. Efficient $^{3}$He/$^{4}$He separation in a nanoporous graphenylene membrane. *Phys. Chem. Chem. Phys.* **2017**, *19*, 21522–21526.

(18) Kou, J.; Zhou, X.; Lu, H.; Wu, F.; Fan, J. Graphyne as the membrane for water desalination. *Nanoscale* **2014**, *6*, 1865–1870.

(19) Kong, X.; Li, L.; Leenaerts, O.; Liu, X. J.; Peeters, F. M. New group-V elemental bilayers: A tunable structure model with four-, six-, and eight-atom rings. *Phys. Rev. B* **2017**, *96*, 035123.

(20) Nulakani, N. V. R.; Subramanian, V. Cp-Graphyne: A Low-Energy Graphyne Polymorph with Double Distorted Dirac Points. *ACS Omega* **2017**, *2*, 6822–6830.

(21) Long, G.; Zhou, Y.; Jin, M.; Kan, B.; Zhao, Y.; Gray-Weale, A.; Jiang, D. E.; Chen, Y.; Zhang, Q. Theoretical investigation on two-dimensional non-traditional carbon materials employing three-membered ring and four-membered ring as building blocks. *Carbon* **2015**, *95*, 1033–1038.

(22) Zhang, L. Z.; Wang, Z. F.; Wang, Z. M.; Du, S. X.; Gao, H. J.; Liu, F. Highly Anisotropic Dirac Fermions in Square Graphynes. *J. Phys. Chem. Lett.* **2015**, *6*, 2959–2962.

(23) Weng, H.; Dai, X.; Fang, Z. Topological semimetals predicted from first-principles Calculations. *J. Phys.: Condens. Matter* **2016**, *28*, 303001.

(24) Jin, Y. J.; Wang, R.; Zhao, J. Z.; Du, Y. P.; Zheng, C. D.; Gan, L. Y.; Liu, J. F.; Xu, H.; Tong, S. Y. The prediction of a family group of two-dimensional node-line semimetals. *Nanoscale* **2017**, *9*, 13112–13118.

(25) Wu, Y.; Wang, L. L.; Mun, E.; Johnson, D. D.; Mou, D.; Huang, L.; Lee, Y.; Bud'ko, S. L.; Canfield, P. C.; Kaminski, A. Dirac node arcs in PtSn$_4$. *Nat. Phys.* **2016**, *12*, 667–671.

(26) Bian, G.; Chang, T. R.; Sankar, R.; Xu, S. Y.; Zheng, H.; Neupert, T.; Chiu, C. K.; Huang, S. M.; Chang, G.; Belopolski, I.; et al. Topological nodal-line fermions in spin-orbit metal PbTaSe$_2$. *Nat. Commun.* **2016**, *7*, 10556.

(27) Schoop, L. M.; Ali, M. N.; Straßer, C.; Topp, A.; Varykhalov, A.; Marchenko, D.; Duppel, V.; Parkin, S. S. P.; Lotsch, B. V.; Ast, C. R. Dirac cone protected by non-symmorphic symmetry and three-dimensional Dirac line node in ZrSiS. *Nat. Commun.* **2016**, *7*, 11696.

(28) Neupane, M.; Belopolski, I.; Hosen, M. M.; Sanchez, D. S.; Sankar, R.; Szlawska, M.; Xu, S. Y.;

Dimitri, K.; Dhakal, N.; Maldonado, P.; et al. Observation of topological nodal fermion semimetal phase in ZrSiS. *Phys. Rev. B* **2016**, *93*, 201104(R).

(29) Bian, G.; Chang, T. R.; Zheng, H.; Velury, S.; Xu, S. Y.; Neupert, T.; Chiu, C. K.; Huang, S. M.; Sanchez, D. S.; Belopolski, I.; et al. Drumhead surface states and topological nodal-line fermions in TlTaSe$_2$. *Phys. Rev. B* **2016**, *93*, 121113(R).

(30) Weng, H.; Liang, Y.; Xu, Q.; Yu, R.; Fang, Z.; Dai, X.; Kawazoe, Y. Topological node-line semimetal in three-dimensional graphene networks. *Phys. Rev. B* **2015**, *92*, 045108.

(31) Xie, L. S.; Schoop, L. M.; Seibel, E. M.; Gibson, Q. D.; Xie, W.; Cava, R. J. A new form of Ca$_3$P$_2$ with a ring of Dirac nodes. *APL Mater.* **2015**, *3*, 083602.

(32) Chan, Y. H.; Chiu, C. K.; Chou, M. Y.; Schnyder, A. P. Ca$_3$P$_2$ and other topological semimetals with line nodes and drumhead surface states. *Phys. Rev. B* **2016**, *93*, 205132.

(33) Zeng, M.; Fang, C.; Chang, G.; Chen, Y. A.; Hsieh, T.; Bansil, A.; Lin, H.; Fu, L. Topological semimetals and topological insulators in rare earth monopnictides. *arXiv* **2015**, 1504.03492v1.

(34) Yu, R.; Weng, H.; Fang, Z.; Dai, X.; Hu, X. Topological Node-Line Semimetal and Dirac Semimetal State in Antiperovskite Cu$_3$PdN. *Phys. Rev. Lett.* **2015**, *115*, 036807.

(35) Wang, J. T.; Weng, H.; Nie, S.; Fang, Z.; Kawazoe, Y.; Chen, C. Body-Centered Orthorhombic C$_{16}$: A Novel Topological Node-Line Semimetal. *Phys. Rev. Lett.* **2016**, *116*, 195501.

(36) Yu, R.; Fang, Z.; Dai, X.; Weng, H. Topological nodal line semimetals predicted from first-principles calculations. *Front. Phys.* **2017**, *12*, 127202.

(37) Feng, B.; Fu, B.; Kasamatsu, S.; Ito, S.; Cheng, P.; Liu, C. C.; Feng, Y.; Wu, S.; Mahatha, S. K.; Sheverdyaeva, P.; et al. Experimental realization of two-dimensional Dirac nodal line fermions in monolayer Cu$_2$Si. *Nat. Commun.* **2017**, *8*, 1007.

(38) Yang, B.; Zhang, X.; Zhao, M. Dirac node lines in two-dimensional Lieb lattices. *Nanoscale* **2017**, *9*, 8740–8746.

(39) Lu, J. L.; Luo, W.; Li, X. Y.; Yang, S. Q.; Cao, J. X.; Gong, X. G.; Xiang, H. J. Two-Dimensional

Node-Line Semimetals in a Honeycomb-Kagome Lattice. *Chin. Phys. Lett.* **2017**, *34*, 057302.

(40) Kresse, G.; Furthmüller, J. Efficient iterative schemes for *ab initio* total-energy calculations using a plane-wave basis set. *Phys. Rev. B* **1996**, *54*, 11169–11186.

(41) Kresse, G.; Joubert, D. From ultrasoft pseudopotentials to the projector augmented-wave method. *Phys. Rev. B* **1999**, *59*, 1758–1775.

(42) Kresse, G.; Hafner, J. *Ab initio* molecular dynamics for open-shell transition metals. *Phys. Rev. B* **1993**, *48*, 13115–13118.

(43) Perdew, J. P.; Burke, K.; Ernzerhof, M. Generalized Gradient Approximation Made Simple. *Phys. Rev. Lett.* **1996**, *77*, 3865–3868.

(44) Togo, A.; Tanaka, I. First principles phonon calculations in materials science. *Scr. Mater.* **2015**, *108*, 1–5.

(45) Baughman, R. H.; Eckhardt, H.; Kertesz, M. Structure-property predictions for new planar forms of carbon: Layered phases containing $sp^2$ and *sp* atoms. *J. Chem. Phys.* **1987**, *87*, 6687–6699.

(46) Yin, W. J.; Xie, Y. E.; Liu, L. M.; Wang, R. Z.; Wei, X. L.; Lau, L.; Zhong, J. X.; Chen, Y. P. R-graphyne: a new two-dimensional carbon allotrope with versatile Dirac-like point in nanoribbons. *J. Mater. Chem. A* **2013**, *1*, 5341–5346.

(47) Zhang, H.; Pan, H.; Zhang, M.; Luo, Y. First-principles prediction of a new planar hydrocarbon material: half-hydrogenated 14,14,14-graphyne. *Phys. Chem. Chem. Phys.* **2016**, *18*, 23954–23960.

(48) Nulakani, N. V. R.; Subramanian, V. A Theoretical Study on the Design, Structure, and Electronic Properties of Novel Forms of Graphynes. *J. Phys. Chem. C* **2016**, *120*, 15153–15161.

(49) Wang, A.; Li, L.; Wang, X.; Bu, H.; Zhao, M. Graphyne-based carbon allotropes with tunable properties: From Dirac fermion to semiconductor. *Diam. Relat. Mater.* **2014**, *41*, 65–72.

(50) Qu, Y.; Li, F.; Zhao, M. Efficient hydrogen isotopologues separation through a tunable potential barrier: The case of a $C_2N$ membrane. *Sci. Rep.* **2017**, *7*, 1483.

(51) Li, F.; Qu, Y.; Zhao, M. Efficient helium separation of graphitic carbon nitride membrane. *Carbon* **2015**, *95*, 51–57.

(52) Biswal, B. P.; Chaudhari, H. D.; Banerjee, R.; Kharul, U. K. Chemically Stable Covalent Organic Framework (COF)-Polybenzimidazole Hybrid Membranes: Enhanced Gas Separation through Pore Modulation. *Chem. Eur. J.* **2016**, *22*, 4695–4699.

(53) Zhang, J.; Wang, R.; Zhu, X.; Pan, A.; Han, C.; Li, X.; Zhao, D.; Ma, C.; Wang, W.; Su, H.; et al. Pseudo-topotactic conversion of carbon nanotubes to T-carbon nanowires under picosecond laser irradiation in methanol. *Nat. Commun.* **2017**, *8*, 683.

(54) Sheng, X. L.; Yan, Q. B.; Ye, F.; Zheng, Q. R.; Su, G. T-Carbon: A Novel Carbon Allotrope. *Phys. Rev. Lett.* **2011**, *106*, 155703.

(55) Heyd, J.; Scuseria, G. E.; Ernzerhof, M. Hybrid functionals based on a screened Coulomb potential. *J. Chem. Phys.* **2003**, *118*, 8207–8215.

(56) Heyd, J.; Scuseria, G. E.; Ernzerhof, M. Erratum: "Hybrid functionals based on a screened Coulomb potential" [J. Chem. Phys. 118, 8207 (2003)]. *J. Chem. Phys.* **2006**, *124*, 219906.

(57) Xu, L. C.; Du, A.; Kou, L. Hydrogenated borophene as a stable two-dimensional Dirac material with an ultrahigh Fermi velocity. *Phys. Chem. Chem. Phys.* **2016**, *18*, 27284–27289.

(58) Li, L.; Kong, X.; Leenaerts, O.; Chen, X.; Sanyal, B.; Peeters, F. M. Carbon-rich carbon nitride monolayers with Dirac cones: Dumbbell $C_4N$. *Carbon* **2017**, *118*, 285–290.

(59) Malko, D.; Neiss, C.; Viñes, F.; Görling, A. Competition for Graphene: Graphynes with Direction-Dependent Dirac Cones. *Phys. Rev. Lett.* **2012**, *108*, 086804.

(60) Li, L.; Leenaerts, O.; Kong, X.; Chen, X.; Zhao, M.; Peeters, F. M. Gallium bismuth halide GaBi-$X_2$ (X = I, Br, Cl) monolayers with distorted hexagonal framework: Novel room-temperature quantum spin Hall insulators. *Nano Res.* **2017**, *10*, 2168–2180.

(61) Kong, X.; Li, L.; Leenaerts, O.; Wang, W.; Liu, X. J.; Peeters, F. M. Quantum anomalous Hall effect in stable 1T-$YN_2$ monolayer with a large nontrivial band gap and high Chern number. *arXiv*

**2017**, 1707.01841v2.

(62) Niu, C.; Buhl, P. M.; Bihlmayer, G.; Wortmann, D.; Dai, Y.; Blügel, S.; Mokrousov, Y. Two-dimensional topological nodal line semimetal in layered $X_2Y$ ($X$ = Ca, Sr, and Ba; $Y$ = As, Sb, and Bi). *Phys. Rev. B* **2017**, *95*, 235138.

(63) Dean, C. R.; Young, A. F.; Meric, I.; Lee, C.; Wang, L.; Sorgenfrei, S.; Watanabe, K.; Taniguchi, T.; Kim, P.; Shepard, K. L.; et al. Boron nitride substrates for high-quality graphene electronics. *Nat. Nanotechnol.* **2010**, *5*, 722–726.

(64) Yang, W.; Chen, G.; Shi, Z.; Liu, C. C.; Zhang, L.; Xie, G.; Cheng, M.; Wang, D.; Yang, R.; Shi, D.; et al. Epitaxial growth of single-domain graphene on hexagonal boron nitride. *Nat. Mater.* **2013**, *12*, 792–797.

(65) Zhao, X.; Li, L.; Zhao, M. Lattice match and lattice mismatch models of graphene on hexagonal boron nitride from first principles. *J. Phys.: Condens. Matter* **2014**, *26*, 095002.

(66) Li, L.; Wang, X.; Zhao, X.; Zhao, M. Moiré superstructures of silicene on hexagonal boron nitride: A first-principles study. *Phys. Lett. A* **2013**, *377*, 2628–2632.

(67) Li, L.; Zhao, M. First-principles identifications of superstructures of germanene on Ag(111) surface and h-BN substrate. *Phys. Chem. Chem. Phys.* **2013**, *15*, 16853–16863.

(68) Grimme, S.; Antony, J.; Ehrlich, S.; Krieg, H. A consistent and accurate *ab initio* parametrization of density functional dispersion correction (DFT-D) for the 94 elements H-Pu. *J. Chem. Phys.* **2010**, *132*, 154104.

(69) Li, L.; Zhao, M. Structures, Energetics, and Electronic Properties of Multifarious Stacking Patterns for High-Buckled and Low-Buckled Silicene on the $MoS_2$ Substrate. *J. Phys. Chem. C* **2014**, *118*, 19129–19138.